\documentclass[aps,prd,twocolumn,floatfix,superscriptaddress,balancelastpage,nofootinbib,amsmath,showpacs]{revtex4}
\usepackage{graphicx}
\usepackage{epsfig}
\usepackage{bm}

\begin{document}

\hspace*{130mm}{\large \tt FERMILAB-PUB-06-480-A}

\title{High-energy neutrinos from astrophysical accelerators of cosmic ray 
nuclei} 

\author{Luis A.~Anchordoqui}
\affiliation{Department of Physics, University of Wisconsin-Milwaukee, 
             P O Box 413, Milwaukee, WI 53201, USA}
\author{Dan Hooper}
\affiliation{Fermi National Accelerator Laboratory, 
             Theoretical Astrophysics, Batavia, IL 60510, USA} 
\author{Subir Sarkar} 
\affiliation{Rudolf Peierls Centre for Theoretical Physics, 
             University of Oxford, 1 Keble Road, Oxford OX1 3NP, UK}
\author{Andrew M. Taylor}
\affiliation{Astrophysics, University of Oxford, 
             Denys Wilkinson Building, Keble Road, Oxford OX1 3RH,
	     UK\\
             now at: Max-Planck-Institut f\"ur Kernphysik, 
             Postfach 103980, D-69029 Heidelberg, GERMANY}
\begin{abstract}
\noindent 
  Ongoing experimental efforts to detect cosmic sources of high energy
  neutrinos are guided by the expectation that astrophysical
  accelerators of cosmic ray protons would also generate neutrinos
  through interactions with ambient matter and/or photons. However
  there will be a reduction in the predicted neutrino flux if cosmic
  ray sources accelerate not only protons but also significant numbers
  of heavier nuclei, as is indicated by recent air shower data. We
  consider plausible extragalactic sources such as active galactic
  nuclei, gamma-ray bursts and starburst galaxies and demand
  consistency with the observed cosmic ray composition and energy
  spectrum at Earth after allowing for propagation through
  intergalactic radiation fields. This allows us to calculate the
  expected neutrino fluxes from the sources, normalized to the
  observed cosmic ray spectrum. We find that the likely signals are
  still within reach of next generation neutrino telescopes such as
  IceCube.
\end{abstract}

\maketitle

\section{Introduction}

It has long been recognised that high energy protons produced in
cosmic ray accelerators would also generate an observable flux of
cosmic neutrinos, mainly through pion production in collisions with
the ambient gas or with radiation fields~\cite{review}. Neutrino
telescopes such as AMANDA/IceCube~\cite{icecube},
ANTARES~\cite{antares}, RICE~\cite{rice}, ANITA~\cite{Barwick:2005hn}
and the Pierre Auger Observatory~\cite{Auger} have presented initial
results and are approaching the level of sensitivity thought to be
required to detect the first high-energy cosmic
neutrinos~\cite{Bednarek:2004ky}.

Candidate sources of high-energy neutrinos are of both Galactic and
extragalactic varieties, the latter being expected to dominate at the
highest energies just as for the parent cosmic rays. Likely Galactic
sources include microquasars~\cite{microquasars} and supernova
remnants~\cite{snr}, while possible extragalactic sources include
active galactic nuclei (AGN)~\cite{agn}, gamma-ray bursts
(GRBs)~\cite{grb}, and starburst galaxies~\cite{starbursts}. If
however the cosmic ray sources accelerate heavy nuclei, the resulting
high energy neutrino spectrum may be altered considerably from the
all-proton picture which is usually assumed. Nuclei undergoing
acceleration can interact with radiation fields in or near the cosmic
ray engine, causing them to photo-disintegrate into their constituent
nucleons which can then proceed to generate neutrinos through
photo-pion interactions. Hence if most of the accelerated nuclei are
broken up into nucleons before they can escape from their sources, the
neutrino spectrum will not differ much from that predicted for proton
accelerators. However if the radiation fields surrounding the sources
are not sufficiently dense to fully disintegrate cosmic ray nuclei,
fewer nucleons will be freed, leading to a reduced neutrino flux. In
fact heavy nuclei can directly photo-produce pions on radiation fields
but since the photo-pion production threshold is much higher than
typical photo-disintegration thresholds, such interactions will be
unimportant except at very high energies, well beyond the
Galactic/extragalactic transition in the cosmic ray spectrum. Of
course both nuclei and nucleons can scatter inelastically with ambient
gas surrounding the sources to produce pions which subsequently decay
into neutrinos.

In this article, we explore the impact of primary nuclei on the
generation of high energy neutrinos in plausible extragalactic sources
of cosmic rays. In Sec.~\ref{wb}, we revisit the ``Waxman-Bahcall
bound'' on the high-energy cosmic neutrino spectrum and generalize
their argument to include cosmic ray nuclei. In Sec.~\ref{specific} we
estimate the photo-nuclear energy loss lengths for three suggested
high-energy cosmic ray accelerators: AGN, GRBs, and starburst
galaxies.  We find that from AGN, nuclei with energies below $\sim
10^{19}$~eV can escape largely intact, while by $10^{19.5}$~eV most
iron nuclei will suffer violently and only nucleons or light nuclei
are expected to escape. In the case of GRB, most energetic nuclei will
undergo complete disintegration and only nucleons can be emitted as
cosmic rays. Starburst galaxies by contrast are ideal candidates for
the emission of ultra-high energy cosmic ray nuclei, as very few
nucleons are dissociated. (Recently, accretion shocks around galaxy
clusters have also been invoked in this context
\cite{clustershocks}.) In Sec.~\ref{implications} we discuss the
implications of these results and derive the neutrino fluxes
associated with the observed ultra-high energy cosmic rays under the
assumption that none of the sources are hidden.  To normalize the
fluxes we require that their energy spectrum and chemical composition
at Earth, after propagation through intergalactic radiation
backgrounds, are consistent with experimental data from Auger and
HiRes. In Sec.~\ref{sensitivity} we show that the expected cosmic
neutrino fluxes are still detectable by kilometer scale neutrino
telescopes, such as IceCube. Our conclusions are presented in
Sec.~\ref{conclusions}.

\section{The Waxman-Bahcall bound}
\label{wb}

Waxman and Bahcall~\cite{wb} pointed out some time ago that an upper
bound can be placed on the diffuse flux of cosmic neutrinos assuming
that they are generated in cosmologically distributed, optically-thin
proton accelerators (see also Refs.~\cite{rpm}
and~\cite{Ahlers:2005sn}). In this Section, we discuss how this bound
is affected if the accelerated particles are nuclei rather than
protons.

First we briefly outline the original argument as applied to proton
accelerators. If the observed flux of ultra-high energy cosmic
rays is the result of cosmologically distributed sources, then the
energy injection rate in the $10^{19} - 10^{21}~{\rm eV})$ energy
range can be inferred to be~\cite{waxman}:
\begin{equation}
\left.E^2_{\rm{CR}} \frac{d\dot{N}_{\rm{CR}}}{dE_{\rm{CR}}}\right|_{E_0} 
= \frac{\dot \epsilon_{\rm CR}^{[10^{19}, 10^{21}]}}{\ln(10^{21}/10^{19})} 
\approx 10^{44}\,\rm{erg}\,\rm{Mpc}^{-3} \rm{yr}^{-1} \,\,,
\end{equation}
where an energy spectrum $\propto E^{-2}$ has been assumed and $E_0 =
10^{19}~{\rm eV}$. The energy density of neutrinos produced through
photo-pion interactions of these protons can be directly tied to the
injection rate of cosmic rays:
\begin{equation}
E^2_{\nu} \frac{dN_{\nu}}{dE_{\nu}}
\approx \frac{3}{8} \epsilon_\pi \, t_{\rm{H}}\,E^2_{\rm{CR}} 
\frac{d\dot{N}_{\rm{CR}}}{dE_{\rm{CR}}},
\end{equation}
where $t_{\rm{H}}$ is the Hubble time and $\epsilon_\pi$ is the
fraction of the energy which is injected in protons lost to photo-pion
interactions.  (The factor of 3/8 comes from the fact that, close to
threshold, roughly half the pions produced are neutral, thus not
generating neutrinos, and one quarter of the energy of charged pion
decays --$\pi^+ \rightarrow \mu^+ \nu_{\mu} \rightarrow e^+ \nu_e
\nu_{\mu} \bar{\nu}_{\mu}$-- going to electrons rather than
neutrinos.) Thus the expected neutrino flux is:
\begin{eqnarray} [E^2_{\nu} \Phi_{\nu}]_{\rm WB} & \approx & (3/8)
  \,\xi_Z\, \epsilon_\pi\, t_{\rm{H}}\, \frac{c}{4\pi}\,E^2_{\rm{CR}}
  \frac{d\dot{N}_{\rm{CR}}}{dE_{\rm{CR}}} \nonumber \\ & \approx & 2.3
  \times 10^{-8}\,\epsilon_\pi\,\xi_Z\, \rm{GeV}\,
  \rm{cm}^{-2}\,\rm{s}^{-1}\,\rm{sr}^{-1},
\label{wbproton}
\end{eqnarray}
where the parameter $\xi_Z$ accounts for the effects of source
evolution with redshift, and is expected to be of order unity. The
``Waxman-Bahcall bound'' is defined by the condition $\epsilon_\pi
=1$.  For interactions with the ambient gas (i.e., $pp$ rather than $p
\gamma$ collisions), the average fraction of the total pion energy
carried by charged pions is $2/3$, compared to $1/2$ in the photo-pion
channel. In this case, the upper bound given in Eq.~(\ref{wbproton})
is enhanced by 33\%~\cite{Anchordoqui:2004bd}.

At production, the neutrino flux consists of equal fractions of
$\nu_e$, $\nu_{\mu}$ and $\bar{\nu}_{\mu}$. Originally, the
Waxman-Bahcall bound was presented for the sum of $\nu_{\mu}$ and
$\bar{\nu}_{\mu}$ (neglecting $\nu_e$), motivated by the fact that
only muon neutrinos are detectable as track events in neutrino
telescopes. Since oscillations in the neutrino sector mix the
different species, we chose instead to discuss the sum of all neutrino
flavors. When the effects of oscillations are accounted for, nearly
equal numbers of the three neutrino flavors are expected at Earth.
Note that IceCube will be capable of detecting and discriminating
between all three flavors of neutrinos~\cite{pakvasalearned}.

Electron antineutrinos can also be produced through neutron
$\beta$-decay. This contribution, however, turns out to be negligible.
To obtain an estimate, we sum over the neutron-emitting sources out to
the edge of the universe at a distance
$\sim 1/H_0$~\cite{Anchordoqui:2003vc}:
\begin{equation}
\Phi_{\overline \nu_e} = \frac{m_n \,\,\xi_Z}{8\,\pi \,\epsilon_0\, H_0}
\int_{\frac{m_n E_{\overline \nu}}{2 \epsilon_0}}^{E_n^{\rm max}}
\frac{dE_n}{E_n} \,\, 
\frac{d\dot{N}_n}{dE_n} \,\, ,
\label{nDK}
\end{equation}
where $d\dot{N}_n/dE_n$ is the neutron volume emissivity and $m_n$ the
neutron mass. Here, we have assumed that the neutrino is produced
monoenergetically in the neutron rest frame, i.e., $\epsilon_0 \sim
\delta m (1 - m_e^2/\delta_m^2)/2 \sim 0.55$~MeV, where $\delta m
\simeq 1.30$~MeV is the neutron-proton mass difference. An upper limit
can be placed on $d\dot{N}_n/dE_n$ by assuming an extreme situation in
which all the cosmic rays escaping the source are neutrons, i.e.,
\begin{equation}
\dot \epsilon_{_{\rm CR}} = \int dE_n\,\, E_n\,\, \frac{d\dot{N}_n}{dE_n}\, .
\end{equation}
With the production rate of ultra-high energy protons $\dot
\epsilon_{_{\rm CR}}^{[10^{19}, 10^{21}]} \sim 5 \times 10^{44}$~erg
Mpc$^{-3}$ yr$^{-1}$~\cite{waxman}, and an assumed injection spectrum
$d\dot{N}_n/dE_n \propto E_n^{-2},$ Eq.~(\ref{nDK}) gives
\begin{equation}
E_{\nu}^2 \Phi_{\overline \nu_e}\approx 3\times 10^{-11}\, \xi_Z \,\,
{\rm GeV} \,  {\rm cm}^{-2} \, {\rm s}^{-1} \,
{\rm sr}^{-1}\,\,,
\end{equation}
which is about three orders of magnitude below the Waxman-Bahcall
bound in Eq.(\ref{wbproton}).

If the injected cosmic rays include nuclei heavier than protons, then
the neutrino flux expected from the cosmic ray sources may be
modified.  Nuclei undergoing acceleration can produce pions, just as
protons do, through interactions with the ambient gas, so the
Waxman-Bahcall argument would be unchanged in this case. However if
interactions with radiation fields dominate over interactions with
matter, the neutrino flux would be suppressed if the cosmic rays are
heavy nuclei. This is because the photo-disintegration of nuclei
dominates over pion production at all but the very highest energies.
Defining $\kappa$ as the fraction of nuclei heavier than protons in
the observed cosmic ray spectrum, the resulting neutrino flux is then
given by:
\begin{equation}
  E^2_{\nu} \Phi_{\nu} \approx  (1 - \kappa) \,\,
[E^2_{\nu} \Phi_{\nu}]_{\rm{WB}} \,\,.
\label{simple}
\end{equation}
The Waxman-Bahcall bound can be turned into a flux prediction by
making further assumptions. For example one can assume that all
charged particles remain trapped within the acceleration region and
only neutrons are able to escape. If this were the case, the energy
fraction of cosmic ray protons lost to photo-pion interactions can be
easily obtained from single $pp$ or $p\gamma$ collisions, yielding
$\epsilon_\pi \sim 0.2 - 0.6.$ This flux estimate, however, does not
account for high energy proton production at the source edge (which would have
a large probability of escaping the acceleration region), threshold
effects (near the photo-pion production threshold, the assumed
relationship between the average energy of the incoming proton and
that of the outgoing neutrino can be signifcantly altered), and energy
degradation of the charged pions propagating through the (possibly
strong) magnetic fields in the plasma.  In the remainder of this
paper, we will estimate the flux of cosmic neutrinos from specific
astrophysical sites, taking into account all of these considerations
and assuming that high energy cosmic rays are constituted of both
protons and nuclei, in conformity with observational data on their
spectrum and composition.

\section{Photo-reaction rates for protons and nuclei: three case examples}
\label{specific}

It is likely that the bulk of the cosmic radiation is created as a
result of some general magneto-hydrodynamic process which channels
kinetic or magnetic energy of cosmic plasmas into charged particles.
The details of the acceleration process and the maximum attainable
energy depend on the time scale over which particles are able to
interact with the plasma.  Sometimes the acceleration region itself
only exists for a limited period of time (e.g. supernovae shock waves
dissipate after about $10^{4}$~yr). If the plasma disturbances persist
for long periods, the maximum energy may be limited by the likelihood
of escape from the region. If one includes the effect of the
characteristic velocity, $\beta c$, of the magnetic scattering
centres, the above argument leads to the so-called ``Hillas
criterion'' for the maximum energy acquired by a charged particle
moving in a medium with magnetic field $B$, $E_{\rm max} \sim 2
\beta\, c\, Ze\,B\, r_{_{\rm L}},$ where $r_{\rm L} \approx Z^{-1}
{(B/\mu{\rm G})}^{-1}\, (E_{\rm CR}/10^{18}~{\rm eV})~{\rm kpc}$ is
the Larmor radius of cosmic rays with charge $Ze$~\cite{Hillas:1984}.
In what follows, we consider the radiation fields associated with
three classes of potential cosmic ray sources and determine the
particle's energy loss rate, assuming that within these sources the
trapping condition for efficient acceleration is fulfilled up to the
highest energies~\cite{Torres:2004hk}.

There are basically four ways in which ultra-high energy cosmic rays
interact with ambient photons: Compton interactions, photo-pair
production in the field of the nucleon/nucleus, photo-disintegration
of the nucleus, and photo-production of hadrons (mainly pions).
Although Compton scattering has no energy threshold, it results in a
negligibly small energy loss rate for high energy cosmic rays, so we
will not consider it further.  Photo-pair and photo-pion production
occur at center-of-mass energies higher than $E_{\rm th}^{e^+ e^-} = 2
m_e \simeq 1$~MeV and $E_{\rm th}^{\pi} = m_\pi (1+m_\pi /2 m_p)
\simeq 145$~MeV, respectively. The relative contributions of these two
processes to the total energy losses is dictated by the ratio of the
product of the corresponding inelasticity and cross-section. For a
relativistic nucleon, this product has an average value of $\sim 5
\times 10^{-31} {\rm cm}^2$ from threshold up to $\sim 500$ MeV in the
case of photo-pair production, and an average value of $\sim 7 \times
10^{-29} {\rm cm}^2$ from threshold up to $\sim 200$~GeV for the case
of photo-pion production~\cite{Begelman}. For nuclei, the energy loss
rate due to photo-pair production is $Z^2/A$ times higher than for a
proton of the same Lorentz factor~\cite{Chodorowski}, whereas the
energy loss rate due to photo-meson production remains roughly the
same (because the cross-section for photo-meson production by nuclei
is proportional to the mass number $A$~\cite{Michalowski:eg}, while
the inelasticity is proportional to $1/A$). However, it is
photo-disintegration rather than photo-pair and photo-meson production
that determines the energetics of ultra-high energy cosmic ray
nuclei~\cite{Puget:1976nz}. Experimental data on photo-nuclear
interactions are consistent with a two-step process: photo-absorption
by the nucleus to form a compound state, followed by a statistical
decay process involving the emission of one or more nucleons. Details
of how the relevant cross sections for all these processes are
calculated and the energy loss rates implemented in our simulation can
be found in Ref.~\cite{prev}.

\subsection{Active Galactic Nuclei}

Among known non-thermal sources in the universe, radio--loud AGN seem
to be the most energetic~\cite{note} hence they have long been
suspected to be likely accelerators of ultra-high energy cosmic
rays~\cite{Biermann:1987ep}. At radio frequencies where very large
baseline interferometers can resolve the emission regions on
milliarcsecond scales, many radio--loud AGN exhibit highly collimated
jets of relativistic plasma with opening angles of a few degrees or
less. AGN come in various guises according to the orientation of their
radio jets and the characteristics of the circum-nuclear matter in
their host galaxies.  The most extreme versions are Fanaroff Riley
radio-galaxies whose radio jet axes are almost in the plane of the
sky, and blazars which have the radio jet axes aligned close to the
line-of-sight to the observer (yielding a significant flux enhancement
through Doppler boosting).

\begin{figure}
\begin{center}
\rotatebox{-90}{\resizebox{6.0cm}{!}
{\includegraphics{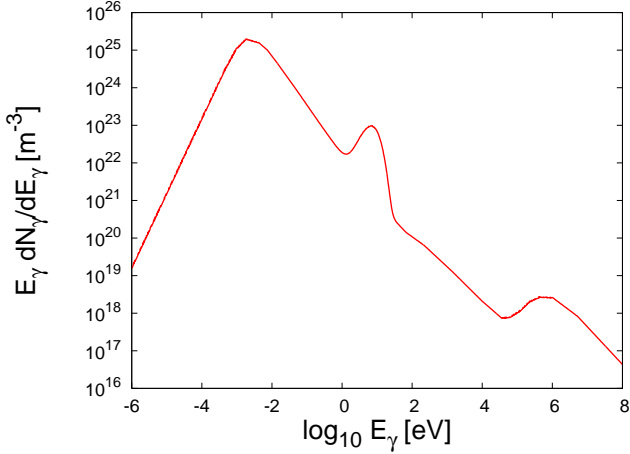}}}
\caption{The photon energy spectrum of a flaring blazar in the
  observer's frame. The normalization is obtained through the relation,
  $\left. U_\gamma/E_\gamma^{\rm peak} = E_\gamma^{\rm peak} {\rm
      d}N/{\rm d}E_\gamma \right|_{\rm peak}$ (N.B.- such a normalisation 
  does not give the photon number density actually measured in the observer's
  frame).}
\label{bkndagn}
\end{center}
\end{figure}

A total of 66 blazars have been detected to date as GeV $\gamma$-ray
sources by EGRET~\cite{Hartman:fc}. In addition, observations from
ground-based air \v{C}erenkov telescopes indicate that in more than
ten of these sources the $\gamma$-ray spectrum extends above a
TeV~\cite{Punch:xw}. The non-thermal emission of these powerful
objects shows a triple-peak structure in the overall spectral energy
distribution. The first component (from radio to $X$-rays) is
generally interpreted as being due to synchrotron radiation from a
population of non-thermal electrons, the second (UV) component as
blackbody radiation originating from the accretion disk, and the third
component ($\gamma$-rays) is explained either as due to inverse
Compton scattering of the same electron population on the various
photon fields traversed by the jet~\cite{Bloom}, or due to development
of electromagnetic cascades triggered by secondary pions (produced
when the highly relativistic baryonic outflow collides with the photon
background~\cite{Mannheim:1993jg} or diffuse gas
targets moving across the jet~\cite{Dar:1996qv}) that cool
instanstaneously via synchrotron radiation. The
characteristic single photon energy in synchrotron radiation
emitted by an electron of energy $E$ is
\begin{equation}
E_\gamma = \left(\frac{3}{2}\right)^{1/2} 
\frac{e\,\hbar c\,E^2\,Bc}{(m_e c^{2})^3} 
\sim 5.4\, B_{\mu{\rm G}}\, E_{19}^{2} \, {\rm TeV}\ \ .\label{synch}
\end{equation}
where, $B_{\mu{\rm G}}$ is the magnetic field in units of $\mu$G and
$E_{19} \equiv E/10^{19}$~eV.  For a proton of the same energy, this
number is $(m_p/m_e)^3 \sim 6 \times 10^9$ times smaller. It is
evident that high energy gamma ray production through proton
synchrotron radiation requires very large ${\cal O}(100\ {\rm G})$
magnetic fields~\cite{Aharonian:2000pv}, and therefore we do not
consider this process in our calculations.

For the background radiation spectrum in AGN, we adopt the B\"ottcher
blazar flaring state model~\cite{Boettcher:1999ab}. The two relevant
components are synchotron radiation peaking at $E_\gamma^{\rm peak}
\sim 0.003~{\rm eV}$, then falling as $dN_{\gamma}/dE_{\gamma} \propto
E_{\gamma}^{-2.3}$, and a $20,000$~K blackbody radiation from the
accretion disk. It is assumed that 10\% of the Eddington luminosity is
radiated as blackbody radiation and 1\% as synchrotron
radiation~\cite{Dermer:1993cz}.

The cosmic ray acceleration process is assumed to occur within a
relativistic blob of plasma moving along the jet with Lorentz factor
$\Gamma \sim 10^{1.5}$. In the rest frame of the plasma, the blob is
assumed to be spherical with radius $R' = c\Delta t' = \Gamma c \Delta
t$, where $\Delta t \approx 10^{4}~{\rm s}$ indicates the typical
duration of a flare. (More exactly we should use the Doppler factor
and average over the angle between the particle velocity and the boost
velocity but ignoring such subtleties does not change our estimates
substantially.) The energy density injected into such a blob
can be estimated from the apparent bolometric luminosity,
$L_{\gamma}$, and the duration of flare
\begin{eqnarray}
 U_{\gamma} &=& \frac{L_{\gamma}\,\, \Delta t}{(4\pi/3)\,\, 
 (c\, \Delta t)^{3}} \nonumber \\
& \approx & 6\times10^{22} \left(\frac{L}{10^{45}~{\rm erg}/{\rm s}}\right)
\left(\frac{\Delta t}{10^{4} {\rm~s}}\right)^{2}~{\rm eV}\, {\rm m}^{-3}. 
\end{eqnarray}
The spectrum of background photons is shown in Fig.~\ref{bkndagn}.
Note that since the blob is relativistic, $L'_{\gamma} \Delta t' =
L_{\gamma}\Delta t/\Gamma$, implying a significant dilution of the
energy density in the plasma rest frame to that expected in a
non-relativistic scenario,
\begin{equation}
U'_{\gamma} = \frac{L'_{\gamma}\Delta t'}{(4/3)\, (c\Delta t')^{3}} = 
\frac{U_\gamma}{\Gamma^{4}} \,.
\end{equation}

\begin{figure}
\begin{center}
\rotatebox{-90}{\resizebox{6.0cm}{!}{\includegraphics{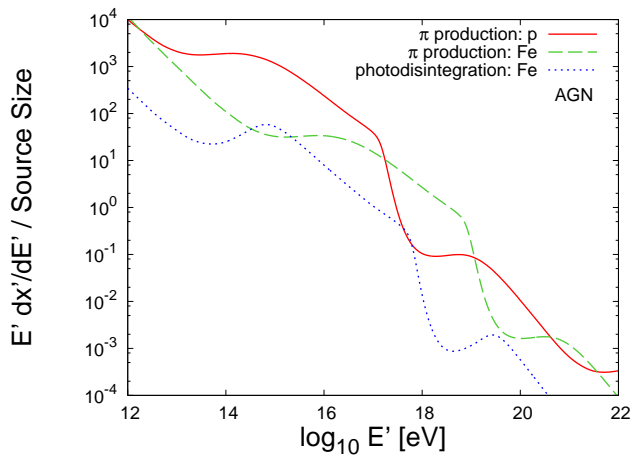}}}
\caption{Comparison of the energy loss lengths for photo-pion production by 
  protons, photo-pion production by iron nuclei and photo-disintegration of
  iron nuclei, as a function of the proton/iron nucleus energy in the
  rest frame of the AGN plasma.  The source size is normalized such
  that $E'{\rm d}x'/{\rm d}E' = K_p (E'_p) /\epsilon_\pi$ at the peak
  energy, which corresponds to a proton energy $E' \equiv E'_p \simeq
  10^{21.2}~{\rm eV}$. The photo-disintegration and photo-pion
  production loss lengths for iron nuclei are normalized using the number
  density of photons given in Fig.~\ref{bkndagn}, reduced by $1/\Gamma^{3}$
  to obtain the value in the plasma rest frame.}
\label{ratesagn}
\end{center}
\end{figure}

The fraction of the total energy in cosmic ray protons, $E'_{p}$,
expected to be lost within the acceleration region to pion production
is just,
\begin{equation}
 \epsilon_{\pi}(E'_{p})
 \approx \frac{R'}{l_{\pi}(E'_{p})}\ ,
\label{emissivity}
\end{equation} 
where
\begin{equation}
 l_{\pi}(E'_{p}) = \frac{E^{\prime 
 {\rm peak}}_\gamma}{K_{p}(E'_{p})U'_{\gamma}\sigma_{\Delta}} = 
 \frac{\Gamma^{4} \, E^{\prime {\rm peak}}_\gamma}{K_{p}(E'_{p})\, \, 
 U_{\gamma}\,\,\sigma_{\Delta}}
\end{equation}
is the proton attenuation length due to interactions with the
radiation field, $\sigma_\Delta$ is the cross-section, and $K(E'_p)$
is the inelasticity of a single collision. (Recall that primed
quantities refer to values measured in the plasma rest frame). 
For protons interacting via the $\Delta$-resonance we find,
\begin{widetext}
\begin{equation}
 \epsilon_{\pi} \approx  \left(\frac{3\times 0.2}{4\pi\Gamma^{2}}\right)
 \left(\frac{L_{\gamma}\Delta t}{E_{\gamma}^{\rm peak}}\right)
 \left(\frac{\sigma_{\Delta}}{(c\Delta t)^{2}}\right) \\
   \approx  553 \left(\frac{10^{1.5}}{\Gamma}\right)^{2}
\left(\frac{L_{\gamma}}{10^{45}~{\rm erg}/{\rm s}}\right)
\left(\frac{3\times10^{-3}~{\rm eV}}{E_{\gamma}^{\rm peak}}
\right)\left(\frac{10^{4}~{\rm s}}{\Delta t}\right) \,,
\end{equation}
\end{widetext}
where the inelasticity, $K_p(E'_p) \approx 0.2$, is kinematically
determined by requiring equal boosts for the decay products of the
$\Delta$~\cite{Stecker:1968uc}. A straightforward calculation shows
that the required proton energy for $\Delta$-resonant production at
the peak of the photon distribution is,
\begin{eqnarray}
E'_{p}=\frac{(m_{\Delta}c^{2})^{2}-(m_{p}c^{2})^{2}}
{4E^{\prime {\rm peak}}_\gamma} \approx  10^{21.2}~{\rm eV},
\end{eqnarray}
where $m_{\Delta} \simeq 1232~{\rm MeV}$ is the mass of the resonance
and $m_{p}$ the proton mass. In Fig.~\ref{ratesagn} we show the
expected photo-pion production and photo-disintegration energy loss lengths 
for protons and nuclei. It is clear that the radiation field becomes thick
to photo-pion interactions at an energy (in the plasma frame) of $E'_p
\approx 10^{18}~{\rm eV}$ (which corresponds to $E_p \approx
10^{19.5}~{\rm eV}$ in the observer's frame), whereas it becomes
opaque to the propagation of nuclei about a decade lower in energy.
One can also verify that photo-disintegration dominates over
photo-pion production by nuclei at all energies, hence neutrino
production through photo-nuclear interactions can be safely neglected.

\subsection{Gamma Ray Bursts}

GRBs are flashes of high energy radiation that can be brighter, during
their brief existence, than any other source in the sky. The bursts
present an amazing variety of temporal profiles, spectra, and
timescales~\cite{Fishman:95}. Our insights into this
phenomenon have been increased dramatically by BATSE observations
of over 2000 GRBs, and more recently, by data from SWIFT~\cite{swift}.

There are several classes of bursts, from single-peaked events,
including the fast rise and exponential decaying (FREDs) and their
inverse (anti-FREDs), to chaotic structures~\cite{Link:1996ss}. There
are well separated episodes of emission, as well as bursts with
extremely complex profiles. Most of the bursts are time asymmetric but
some are symmetric. Burst timescales range from about 30~ms to
several minutes.

The GRB angular distribution appears to be isotropic, suggesting a
cosmological origin~\cite{Meegan:xg}. Furthermore, the detection of
``afterglows'' --- delayed low energy (radio to $X$-ray) emission ---
from GRBs has confirmed this via the redshift determination of several
GRB host-galaxies~\cite{Metzger:1997wp}.

If the sources are so distant, the energy necessary to produce the
observed events by an intrinsic mechanism is astonishing: about
$10^{51}$ erg of gamma rays must be released in less than 1 second.
The most popular interpretation of the GRB-phenomenology is that the
observable effects are due to the dissipation of the kinetic energy of
a relativistic expanding plasma wind, a ``fireball''~\cite{Piran:kx}.
Although the primal cause of these events is not fully understood, it
is generally believed to be associated with the core collapse of
massive stars (in the case of long duration GRBs) and stellar collapse
induced through accretion or a merger (short duration
GRBs)~\cite{meszaros}.

The very short timescale observed in the light curves indicate an
extreme compactness that implies a source which is initially opaque
(because of $\gamma \gamma $ pair creation) to $\gamma$-rays.  The
radiation pressure on the optically thick source drives relativistic
expansion, converting internal energy into the kinetic energy of the
inflating shell. Baryonic pollution in this expanding flow can trap
the radiation until most of the initial energy has gone into bulk
motion with Lorentz factors of $\Gamma \sim 10^2 -
10^3$~\cite{Waxman:2001tk}.  (In our calculations we set $\Gamma =
10^{2.5}$). The kinetic energy can be partially converted into heat
when the shell collides with the interstellar medium or when shocks
within the expanding source collide with one another. The randomized
energy can then be radiated by synchrotron radiation and inverse
Compton scattering yielding non-thermal bursts with timescales of
seconds. Charged particles may be efficiently accelerated to
ultra-high energies in the fireball's internal shocks, hence GRBs are
often considered as potential sources of cosmic
rays~\cite{Waxman:1995vg}.

\begin{figure}
\begin{center}
\rotatebox{-90}{\resizebox{6.0cm}{!}
{\includegraphics{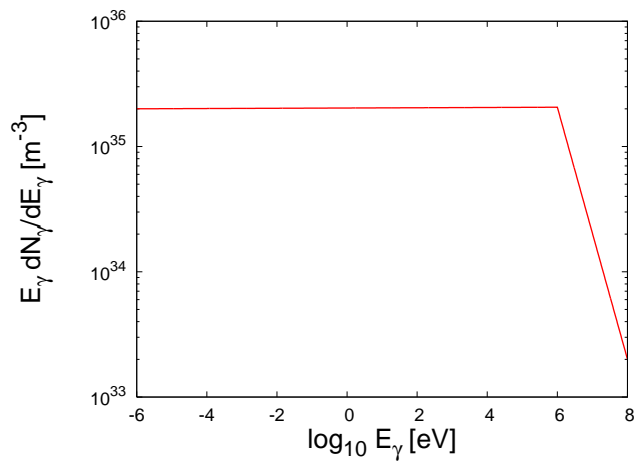}}}
\caption{The photon energy spectrum of a GRB in the observer's
  frame. The normalization is obtained from the condition $\left.
    U_\gamma/E_\gamma^{\rm break} = E_\gamma^{\rm break} {\rm d}N/{\rm
      d}E_\gamma \right|_{\rm break}$ (N.B.- such a normalisation 
  does not give the photon number density actually measured in the 
  observer's frame)}
\label{bkndgrb}
\end{center}
\end{figure}

\begin{figure}
\begin{center}
\rotatebox{-90}{\resizebox{6.0cm}{!}{\includegraphics{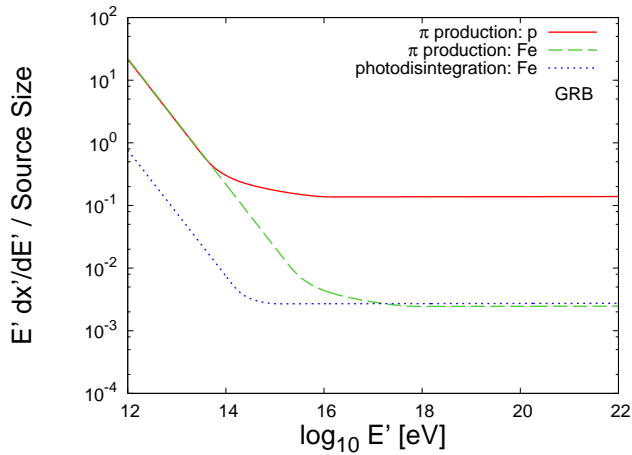}}}
\caption{Comparison of the photo-pion production and
  photo-disintegration energy loss lengths as a function of the
  proton/iron nucleus energy in the rest frame of a GRB fireball. The
  source size is normalized such that 20\% of the accelerated protons
  at a laboratory (observer's) energy of $10^{16}$~eV are converted to
  pions.}
\label{ratesgrb}
\end{center}
\end{figure}

To describe the radiation fields associated with GRBs, we adopt a
standard broken power-law spectrum: ${\rm d}N_{\gamma}/{\rm
  d}E_{\gamma} \propto E_{\gamma}^{-\beta}$, where $\beta$ = 1, 2
respectively at energies below and above the break energy,
$E_\gamma^{\rm break} = 1~{\rm MeV}$~\cite{grb}, which fits the BATSE
data well~\cite{Band}. In many ways the situation is similar to the
blob of emitted plasma in the AGN model. However, in the fireball's
{\em comoving} frame, a spherical shock expands relativistically in
all directions (with Lorentz factor $\Gamma$), and thus a change in
geometry is required.  In a frame moving with Lorentz factor $\Gamma$
towards the observer, the shock has thickness $\Delta R'$ (=$R'/\Gamma$), 
where $R'$ is
the initial size of the compact object before the fireball phase. 
In the observer's frame, the shock is further compressed into a thin
shell of thickness $\Delta R'/\Gamma$ (=$R'$/$\Gamma^2$). 
Therefore, the fiducial value for the energy density, $U'_{\gamma}$, of a 
shell of luminosity, $L'_{\gamma}$, radius $R'$, and thickness $\Delta R'$
is~\cite{Halzen:1998mb},
\begin{widetext}
\begin{equation}
 U'_{\gamma} = \frac{L'_{\gamma}\,\,\Delta t'}{4\pi\, R'^{2}\,\Delta R'} 
 = \frac{L_{\gamma}\Delta t}{\Gamma^{6}4\pi (c\Delta t)^{3}} 
 = 2\times10^{26}\left(\frac{10^{2.5}}{\Gamma}\right)^{6}
 \left(\frac{L_{\gamma}}{10^{51}{\rm~erg~s^{-1}}}\right)
 \left(\frac{2 \times 10^{-3}{\rm~s}}{\Delta t}\right)^{2} {\rm~eV~m^{-3}}
\end{equation}
where $R' = \Gamma^{2}\,c\Delta t$ and $\Delta R' = \Gamma\,c\Delta t$. 
The spectrum of this radiation field is shown in Fig.~\ref{bkndgrb}.

The fraction of energy deposited in the GRB by an ultra-relativistic
proton of energy $E'_{p}$ is,
\begin{equation}
 \epsilon_{\pi}(E'_{p}) \approx \frac{\Delta R'}{l_{\pi}(E'_{p})} \approx \left(\frac{K_{p}(E'_{p})}{4\pi\Gamma^{5}}
 \right)\left(\frac{L_{\gamma}\Delta t}{E_{\gamma}^{\prime {\rm break}}}
 \right)\left(\frac{\sigma_{\Delta}}{(c\Delta t)^{2}}\right) \, .
\label{epsilongrb}
\end{equation}
For protons interacting through the $\Delta$-resonance with photons of
energy $E_\gamma^{\prime {\rm break}} \simeq 10 {\rm keV}$,
Eq.~(\ref{epsilongrb}) yields:
\begin{equation}
 \epsilon_{\pi} \approx \left(\frac{0.2}{4\pi\Gamma^{4}}
 \right)\left(\frac{L_{\gamma}\Delta t}{E_{\gamma}^{\rm break}}\right)
 \left(\frac{\sigma_{\Delta}}{(c\Delta t)^{2}}\right)
 = 0.2 \left(\frac{10^{2.5}}{\Gamma}\right)^{4}\left(\frac{L_{\gamma}}
 {10^{51} {\rm~erg~s^{-1}}}\right)\left(\frac{1{\rm~MeV}}
 {E_{\gamma}^{\rm break}}\right)
 \left(\frac{2 \times 10^{-3}{\rm~s}}{\Delta t}\right) \,\, .
\end{equation}
\end{widetext}
In Fig.~\ref{ratesgrb} we show the expected photo-pion production and
photo-disintegration energy loss length for protons and nuclei. This figure
highlights that (with our assumptions) GRBs are expected to be
optically thin to $p\gamma$ interactions below $10^{16}$~eV, with
protons undergoing at most one interaction before leaving the source
region. It is worthwhile pointing out that although the nucleus
photo-disintegration and photo-pion production loss lengths become comparable
at laboratory (observer) energies $E > 10^{16}~{\rm eV}$, most of the
charged pions produced in this energy regime readily lose energy
through synchroton radiation in the strong $\vec B$-field of the
fireball. For example, at $10^{17}~{\rm eV}$ only 1\% of the produced
pions decay before losing a significant fraction of their energy.
Since the neutrino flux falls by more than two orders of magnitude per
decade increase in energy, the contribution from pions suffering
energy degradation would be negligible, and therefore can be ignored.

\subsection{Starbursts}

Starbursts, galaxies undergoing an episode of large-scale star
formation, have also been proposed as sources of ultra-high energy
cosmic rays~\cite{Elbert:1994zv}. These environments feature strong
infrared emission by dust associated with high levels of interstellar
extinction, strong UV spectra from the Lyman $\alpha$ emission of hot
OB stars, and considerable radio emission produced by recent supernova
remnants. The central regions of starburst galaxies can be orders of
magnitude brighter than those of normal spiral galaxies. From such an
active region, a galactic-scale superwind (driven by the collective
effect of supernovae and winds from massive stars) can conceivably
accelerate cosmic ray nuclei to ultra-high energies.

The Lyman-$\alpha$ background is powered by the rich OB and red
supergiant stellar populations within the inner core of these
galaxies, which typically has a radius $R \sim 100~{\rm pc}$. The
average density in the region depends on the temperatures $T_{\rm
  OB}$ and $T_{\rm SG}$ of O, B, and red supergiant
stars, respectively, and the dilution due to the inverse square law.
Specifically, for a region with $N_{\rm OB}$ OB stars and $N_{\rm SG}$
red supergiant stars, the photon density was estimated to
be~\cite{Anchordoqui:2006pd}
\begin{equation}
n = \frac{9}{4} \
\left[\frac{n^{\rm BE}_{T_{\rm OB}}(\epsilon) \,
N_{\rm OB} \,R_{\rm OB}^2 + n^{\rm BE}_{T_{\rm SG}}(\epsilon) \,
N_{\rm SG} \,R_{\rm SG}^2}{R^2} \right],
\label{Jack}
\end{equation}
where 
\begin{equation}
n^{\rm BE}_T (\epsilon) = (\epsilon/\pi)^2\
\left[e^{\epsilon/T}-1 \right]^{-1} 
\label{nBE}
\end{equation}
is a Bose-Einstein distribution with energy $\epsilon$ and temperature
$T$ (normalized so that the total number of photons in a box is $\int
n (\epsilon) d\epsilon$). Here, $R_{\rm OB (SG)}$ is the radius of the
OB (SG) stars and the factor 9/4 comes from averaging the inverse
squares of the distance of an observer to uniformly distributed
sources in a spherical region of radius $R$~\cite{note94}. We assume
the OB and red supergiant stars to constitute 90\% and 10\%,
respectively, of the total population of 27,000~\cite{Forbes}.

\begin{figure}
\begin{center}
\rotatebox{-90}{\resizebox{6.0cm}{!}
{\includegraphics{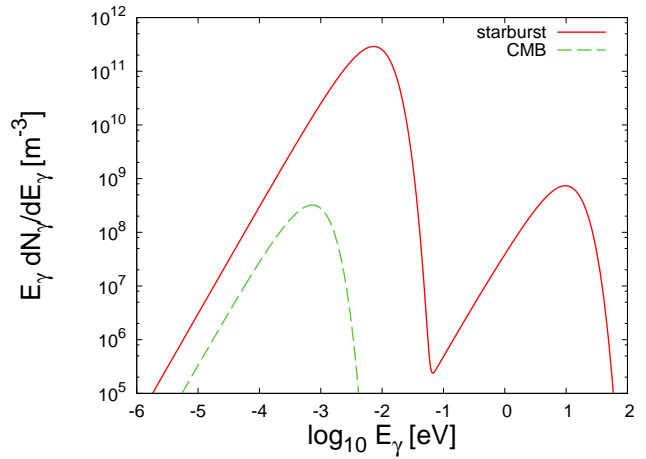}}}
\caption{Thermal photon spectrum of starburt galaxies as derived from
  Eq.~(\ref{Jack}). We have taken for the OB stars a surface
  temperature $T_{\rm OB} = 40,000$~K and radius $R_{\rm 0B} = 15
  R_\odot$, and for the cooler red supergiants, $T_{\rm SG} = 4,000$~K
  and $R_{\rm SG} = 50 \odot$. The spectrum of the cosmic microwave
  background is shown for comparison.}
\label{bkndS}
\end{center}
\end{figure}

We also adopt a single component dust model in which clumpy dust
surrounds the starburst region, in thermal equilibrium with it and
heated to a temperature of 30~K~\cite{Chanial:2006kd}. About 90$\%$ of
the stellar light is assumed to be reprocessed into IR radiation and
the resulting thermal photon spectrum is shown in Fig.~\ref{bkndS}.

Iron nuclei with Lorentz factors up to $\sim 10^6$ may be accelerated
in supernova explosions~\cite{Cesarsky}. Despite the starburst region
being only about $\sim 100~{\rm pc}$ in size, most of these nuclei may
be trapped there through diffusion in milli-Gauss magnetic
fields~\cite{Lai:2003vr} for $\sim 10^4~{\rm
  yr}$~\cite{Bednarek:2003cx}. However since magnetic rigidity
increases with Lorentz factor, the diffusion time would decrease with
rising energy.  In our analysis we adopt the cosmic ray time
delay
\begin{equation}
\tau_{\rm delay}= 400 \, 
Z \left(\frac{E}{5\times 10^{15}~{\rm eV}}\right)^{-1/3}~{\rm yr} \,\, ,
\end{equation}
as expected due to trapping in a Kolmogorov spectrum of magnetic field
fluctuations~\cite{Blasi:1998xp}.

\begin{figure}
\begin{center}
\rotatebox{-90}{\resizebox{6.0cm}{!}{\includegraphics
{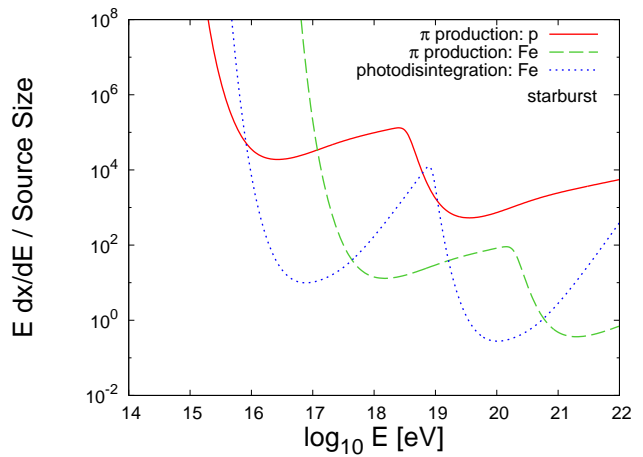}}}
\caption{Comparison of the photo-pion and photo-disintegration
  energy loss lengths as a function of the proton/iron nucleus energy in
  a starburst galaxy.}
\label{ratesS}
\end{center}
\end{figure}

Combining the above estimates of the propagation time, the size of the
starburst region and the number density of UV and IR photons, we find
the proton and nucleus photo-production energy loss lengths shown in
Fig.~\ref{ratesS}. It is seen that the medium is (almost) transparent
to the propagation of cosmic rays. Photo-pion production by nuclei
will not be important until very high energies (Lorentz factors
$\gg 10^{10}$), hence can be neglected.

\begin{figure}
\begin{center}
\rotatebox{-90}{\resizebox{6.0cm}{!}{\includegraphics
{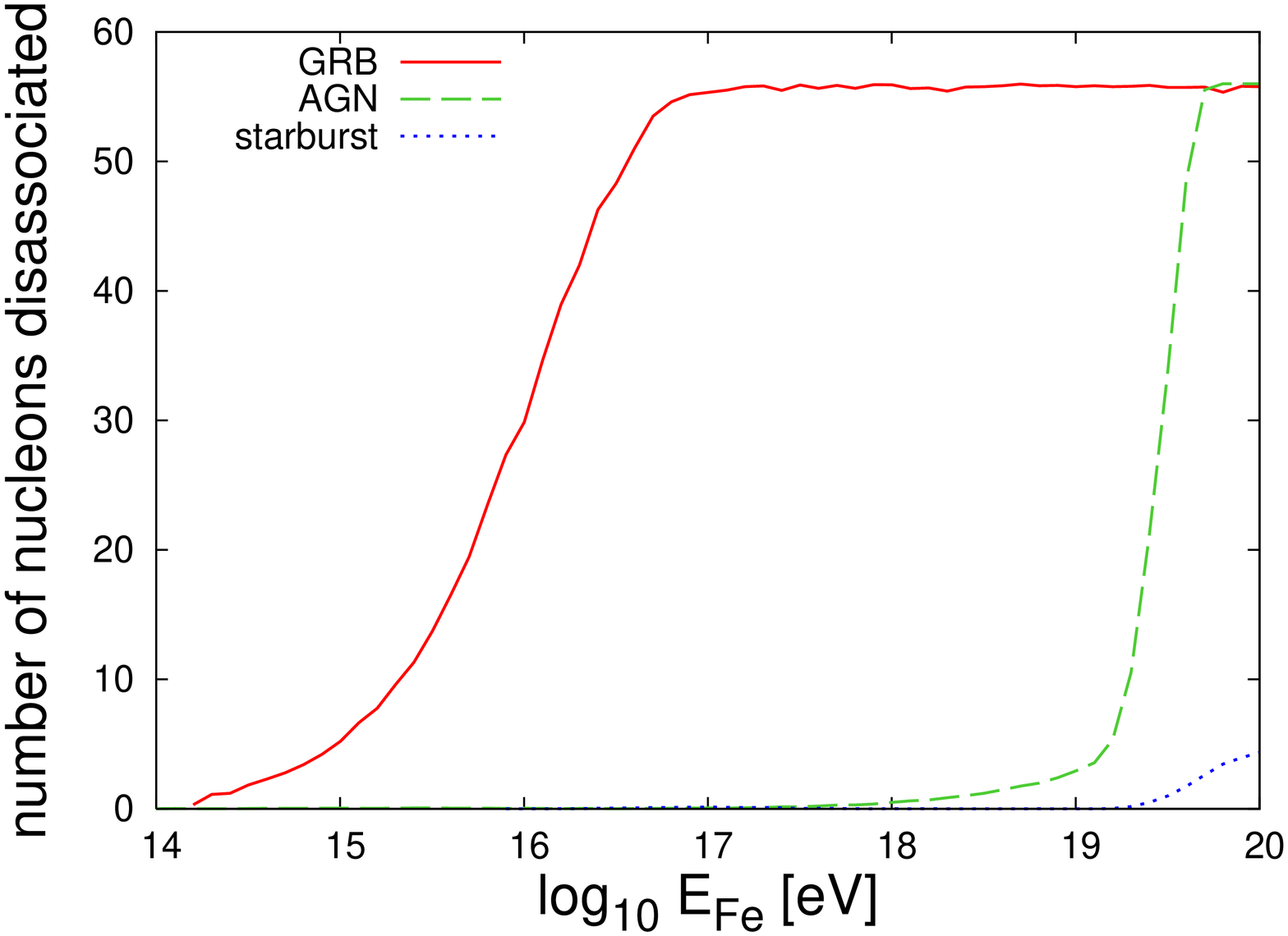}}}\\ 
\vspace{5mm}
\rotatebox{-90}{\resizebox{6.0cm}{!}{\includegraphics{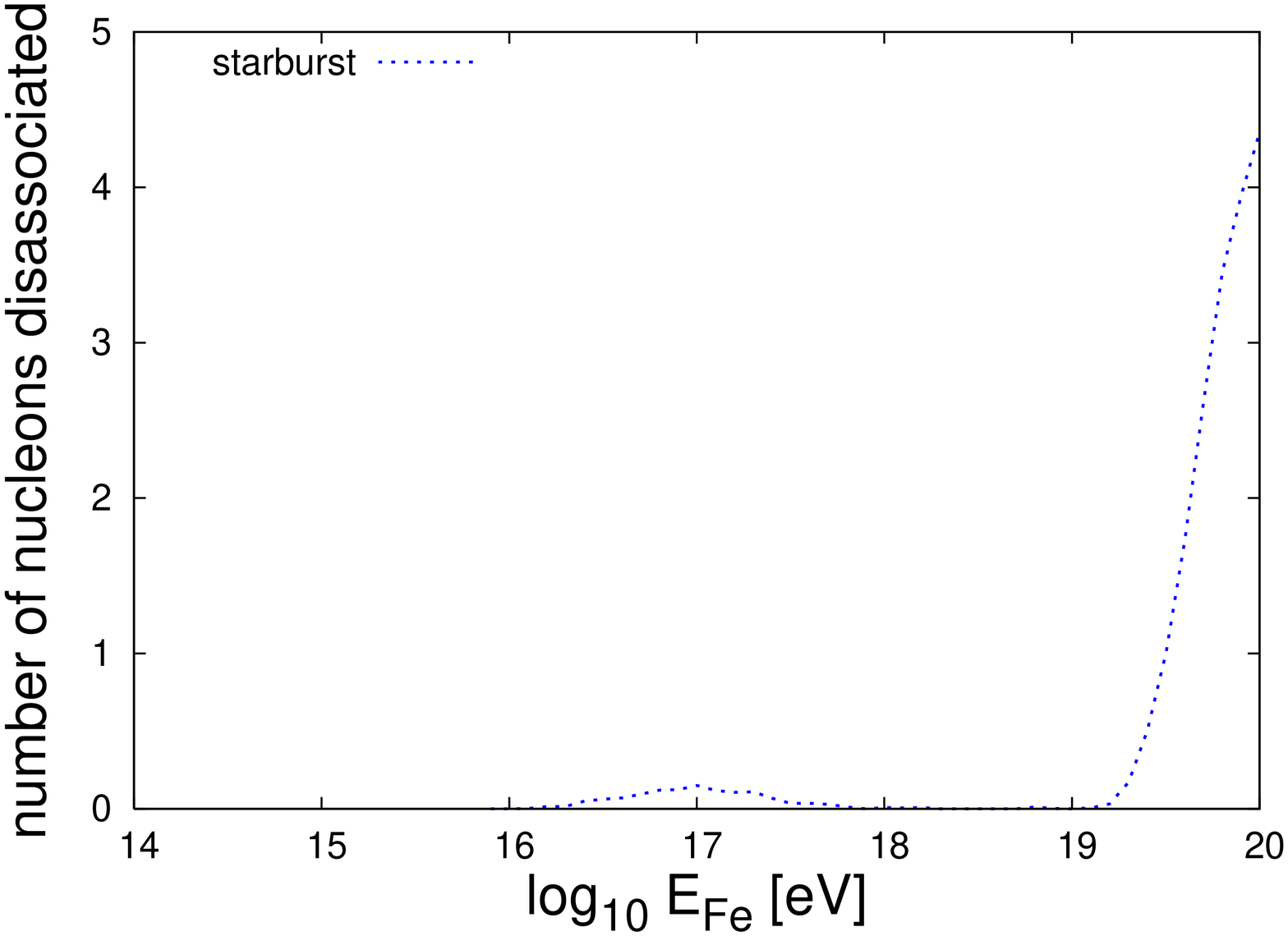}}}
\caption{The top panel shows the average number of nucleons
  photo-dissociated from an iron nucleus of a given energy for the
  radiation field spectra of AGN, GRBs, and starbursts. The lower
  panel shows the starburst case in more detail.}
\label{chain}
\end{center}
\end{figure}

Before proceeding further, it is important to note that pions produced
through hadronic interaction with the ambient gas would also
contribute to the flux of neutrinos. The exact intensity of this
``$pp$ hadronic component'' is currently under
debate~\cite{starbursts,Stecker:2006vz} but an upper bound can
certainly be placed from the non-observation of $\gamma$-rays from
nearby starburst galaxies~\cite{Itoh:2002vf}. In this regard, the
improved sensitivity of GLAST over EGRET is significant --- detection of
even a single starburst galaxy by GLAST would imply that such objects
make a considerable contribution to the diffuse flux of extra-galactic
neutrinos~\cite{Thompson:2006np}. To be conservative in our
calculations we do not consider contributions from the $pp$ channel.

{\em Our essential results so far can be summarized as follows with reference 
to Fig.~\ref{chain}):}\\
{\em (1) In AGN, nuclei with energies below about $10^{19}$~eV can escape
largely intact, while more energetic nuclei undergo an increasing
degree of disintegration; by $10^{19.5}$~eV, most iron nuclei from AGN
will suffer violently, and only nucleons or very light nuclei are
expected to escape.}\\
{\em (2) In GRB, most nuclei will undergo complete disintegration
and only nucleons can be emitted as cosmic rays.}\\
{\em (3) Starburst galaxies are ideal candidates for the emission of
ultra-high energy cosmic ray nuclei, as very few nucleons are
dissociated even at the highest energies.}\\

We can now calculate the neutrino flux associated with the observed
ultra-high energy cosmic rays, taking into account observational data
on their energy spectrum and showering characteristics which suggest
that they consist of a {\em mixture of protons and heavy nuclei},
rather than being purely protons. In other respects we make the same
assumptions as Waxman and Bahcall~\cite{wb}, e.g. that the
extragalactic sources are not hidden.

\section{The cosmic neutrino flux from extragalctic sources}
\label{implications}

As we have seen above, cosmic ray nuclei which escape from the
acceleration region unscathed do not contribute to the source neutrino
spectrum. Therefore, in order to estimate the latter, we first need to
determine the fraction of heavy nuclei in the high energy cosmic rays
arriving at Earth, which we denote by $\kappa$.

This issue is closely connected with that of the ``cross-over'' energy
at which a transition occurs between Galactic and extragalactic cosmic
rays. It would be natural to expect a flattening of the energy
spectrum at this point as the harder subdominant extragalactic
component takes over from the softer Galactic component. Such an
``ankle'' was indeed observed by Fly's Eye~\cite{Bird:1993yi} at $\sim
10^{18.5 - 18.7}$~eV which is roughly the energy at which the (proton)
Larmor radius begins to exceed the thickness of the Milky Way
disk and one expects the Galactic component of the spectrum to die
out. The end-point of the Galactic flux ought to be dominated by heavy
nuclei, as these have a smaller Larmor radius for a given energy, and
the data is indeed consistent with a transition from heavy nuclei to
a lighter composition at the ankle.

However recent HiRes data indicate that this change in the cosmic ray
composition occurs at a much lower energy of $\sim
10^{17.6}$~eV~\cite{Bergman:2004bk}, where the spectral slope steepens
from $E^{-3}$ to $E^{-3.3}$~\cite{Abu-Zayyad:2000ay}. This ``second
knee'' in the spectrum, recognised originally in AGASA
data~\cite{Nagano:1991jz}, can be explained~\cite{Berezinsky:2002nc}
as arising from energy losses of extra-galactic protons over cosmic
distances due to $e^+ e^-$ pair-production on the cosmic microwave
background (CMB). The ``ankle'' is now interpreted as the minimum in the
$e^+ e^-$ energy-loss feature~\cite{Berezinsky:2005cq} and this
requires that there be an almost total absence of iron nuclei ($\alt
0.05\%$ for $E>10^{18}~{\rm eV}$) in extragalactic cosmic
rays~\cite{Berezinsky:2004wx}.

\begin{figure}
\begin{center}
\rotatebox{-90}{\resizebox{6.0cm}{!}{\includegraphics{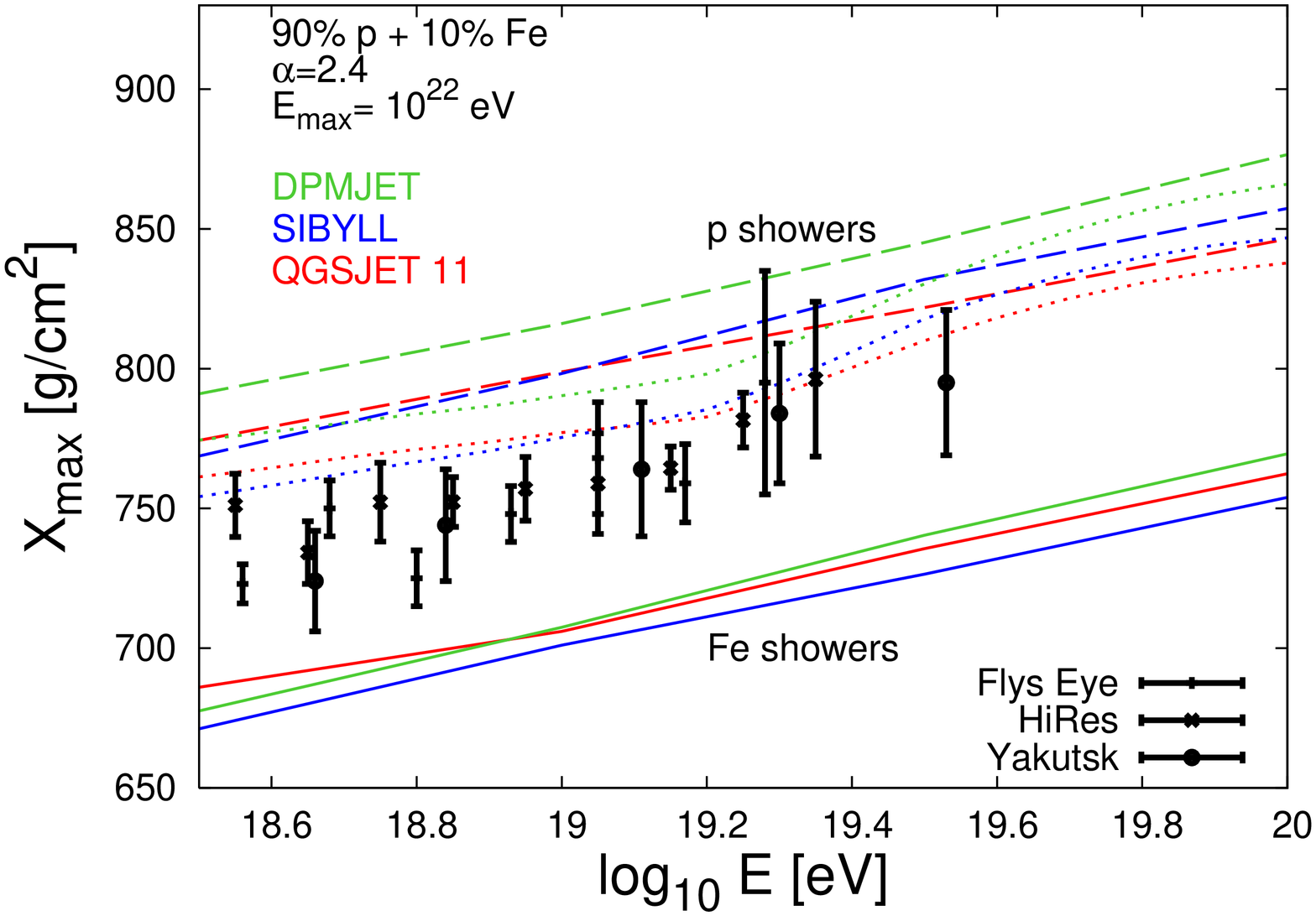}}}\\
\rotatebox{-90}{\resizebox{6.0cm}{!}{\includegraphics{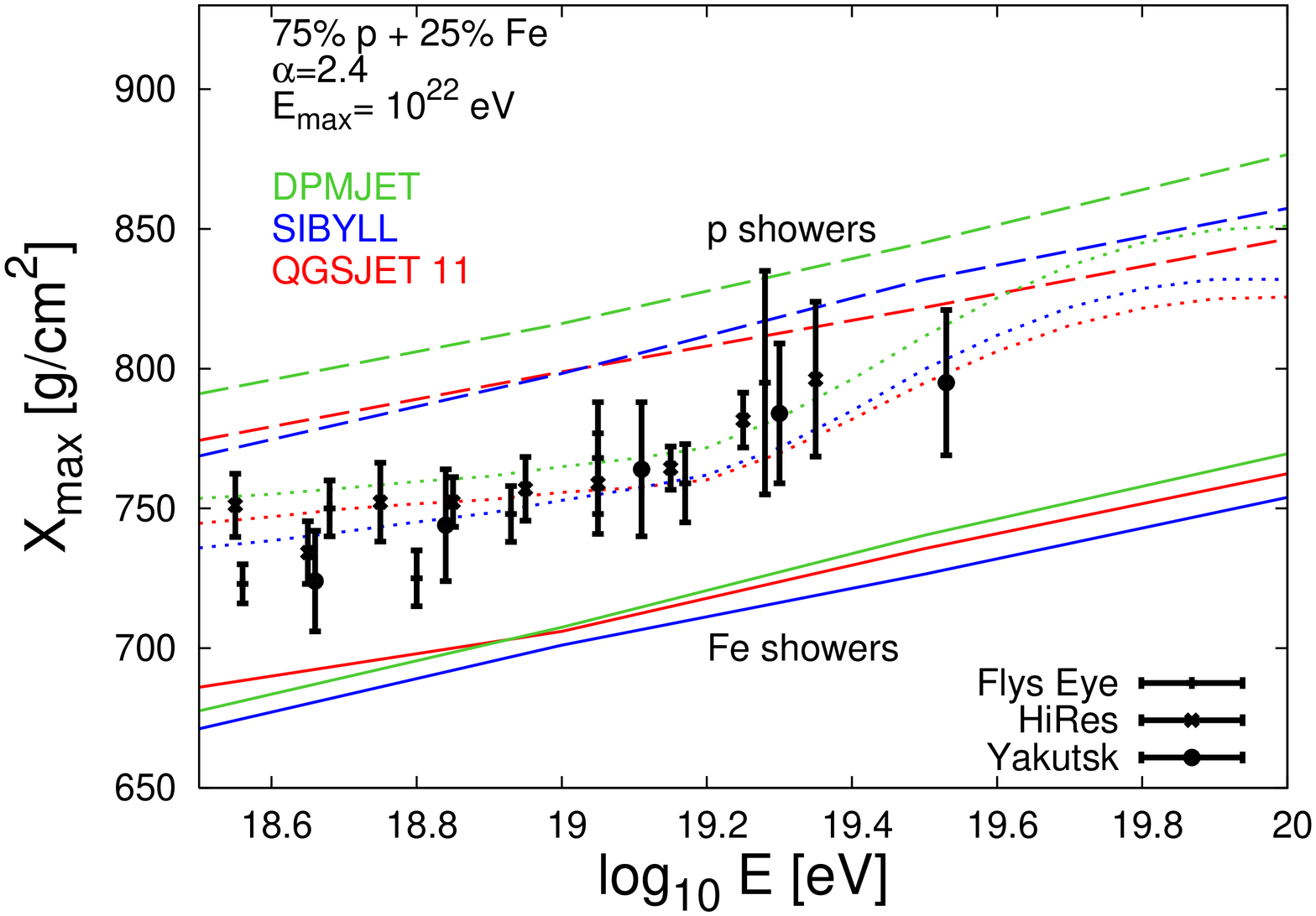}}}
\caption{The top (bottom) panel shows the depth of shower maximum,
  $X_{\rm max}$, for cosmic rays with energy spectrum $\propto
  E^{-\alpha}$ which are a mixture of 90\% (75\%) protons and 10\%
  (25\%) iron. The solid, dashed, and dotted lines are the predicted
  values of $X_{\rm max}$ for a pure iron reaching the Earth's
  atmosphere, pure proton, and the mixed composition (at the source),
  respectively. The spread in the predictions for different hadronic
  interaction models is also shown. The experimental data are from the
  Fly's Eye~\cite{Bird:1993yi}, HiRes~\cite{Abbasi:2004nz} and
  Yakutsk~\cite{yakutsk} experiments.}
\label{xmax}
\end{center}
\end{figure}

However the data on the composition inferred from different analyses
of the characteristics of cosmic ray air showers do not support such a
simple picture \cite{Anchordoqui:2004xb}. For example the depth of
shower maximum $X_{\rm max}$ at $E>10^{18}~{\rm eV}$ is {\em not}
consistent with a pure proton composition, even allowing for the
uncertainties in modelling hadronic interactions at such high enegies.
This is seen in Fig.~\ref{xmax} where we show the evolution of $X_{\rm
  max}$ for both proton and iron showers as obtained from extensive
air shower simulations using the program CORSIKA (version
6.20)~\cite{Knapp} along with experimental data from Fly's Eye, HiRes
and Yakutsk. The predictions of three different hadronic interaction
models (DPMJET~\cite{Ranft:1994fd}, SIBYLL~\cite{Fletcher:1994bd}, and
QGSJET II~\cite{Kalmykov:1997te}) are shown so that an impression of
the modelling uncertainties may be gained from the spread between
them.

In fact, recent analyses~\cite{Allard:2005ha,prev} show that the
$X_{\rm max}$ data, as well as the data on the energy spectrum from
both HiRes~\cite{Abu-Zayyad:2002sf} and the Pierre Auger
Observatory~\cite{Sommers:2005vs}, can {\em simultaneously} be
reproduced if the extra-galactic cosmic rays contain a substantial
fraction of heavy nuclei, and the Galactic/extra-galactic transition
occurs, as was believed originally, at the ``ankle'' in the spectrum.

\begin{figure}
\begin{center}
\rotatebox{-90}{\resizebox{6.0cm}{!}
{\includegraphics{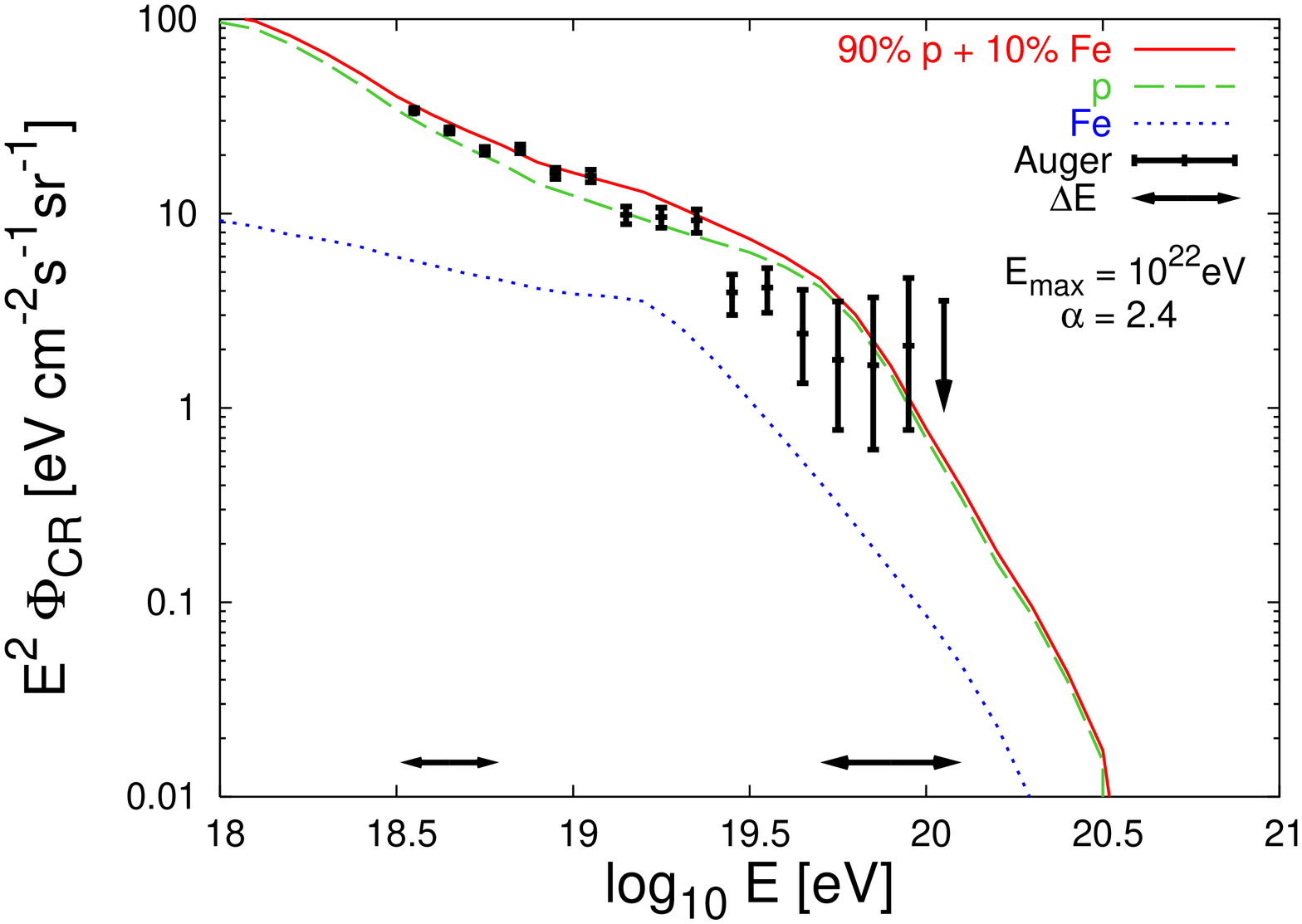}}}\\
\rotatebox{-90}{\resizebox{6.0cm}{!}
{\includegraphics{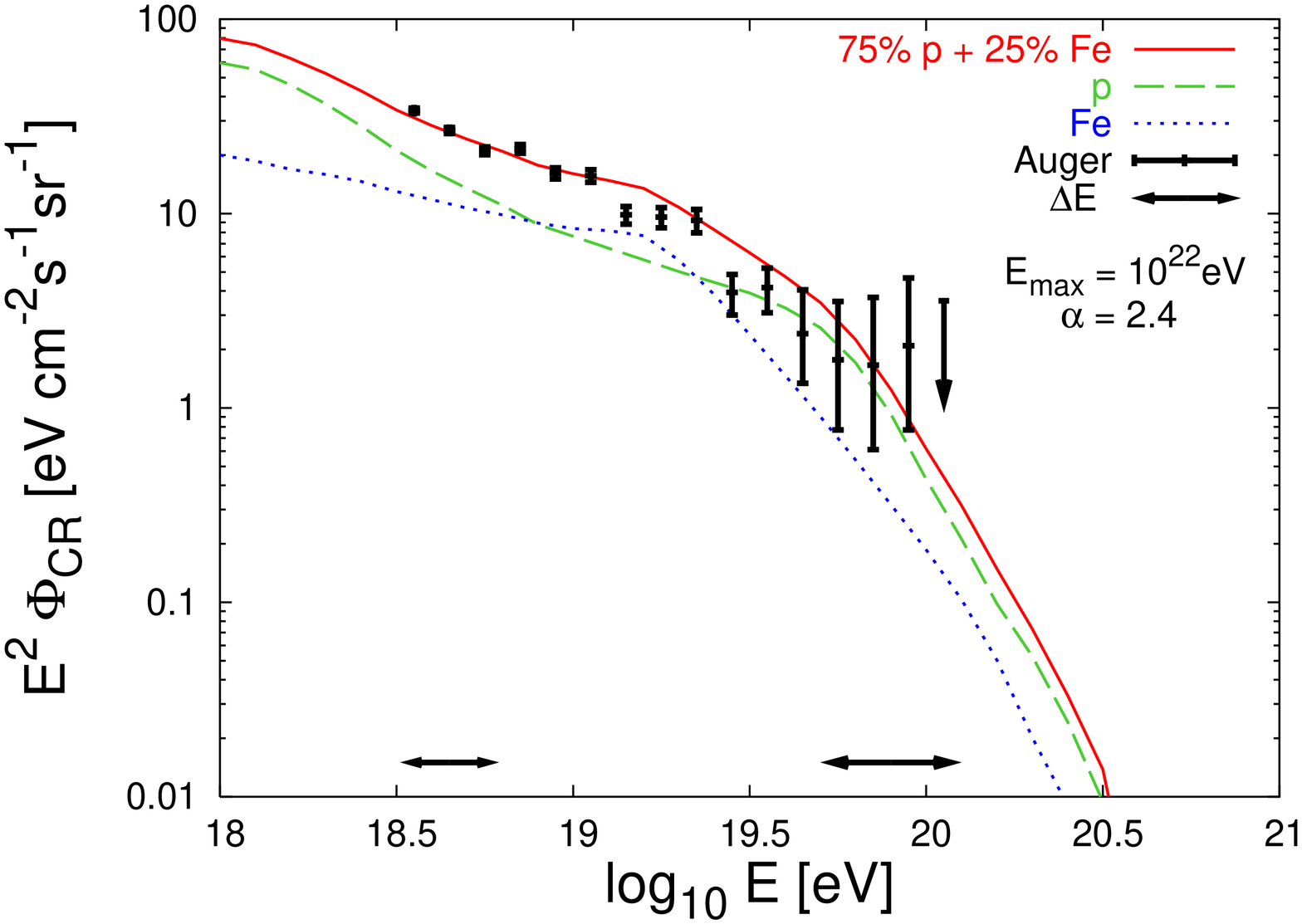}}}
\caption{Best fit of the Auger energy spectrum~\cite{Sommers:2005vs},
  assuming a constant comoving density of sources which emit both
  protons and iron nuclei with energy spectrum $\propto E^{-\alpha}$.
  In the fitting procedure we used data above $E_- = 10^{18.5}$~eV
  (top panel) and above $E_-=10^{18.7}$~eV (lower panel), finding
  $\chi^2/{\rm d.o.f.}  = 52.5/15$ and $\chi^2/{\rm d.o.f.} =
  41.6/13$, respectively. The favoured mixture is 10\% (25\%) iron and
  90\% (75\%) for $E_- = 10^{18.5}$~eV ($E_- = 10^{18.7}$~eV) --- the
  dependence on $E_-$ arises because lower energy points have smaller
  statistical uncertainties. The horizontal error bars indicate the
  systematic uncertainties in energy measurement.}
\label{fit}
\end{center}
\end{figure}

In previous work \cite{prev} we undertook a detailed calculation of
the intergalactic propagation of ultra-high energy heavy nuclei
through known cosmic radiation fields, in order to find the energy
spectrum and composition at Earth. To keep things simple we now
consider only protons and iron nuclei. Armed with our previous
results, we estimate the fraction of iron nuclei, $\kappa$, in
extragalactic cosmic rays, by finding the linear combination of pure
iron and pure proton energy spectra that matches the Auger
data~\cite{Sommers:2005vs} best, after propagation effects are
accounted for. Our results, shown in Fig.~\ref{fit}, indicate that a
mixed composition with 75\% protons and 25\% iron nuclei best
reproduces the data (although the fit quality is poor, suggesting
correlated and/or overestimated uncertainties). As seen in
Fig.~\ref{xmax}, this particular mixed composition at source also
allows a good match to the $X_{\rm max}$ data. A better fit can
perhaps be obtained by considering a more complex composition (e.g.
reflecting the composition of matter in plausible sources) but the
quality of present data is not good enough to warrant this. Even in
this simple 2-component model, we obtain a distinct improvement in the
fit by increasing $\kappa$ from 10\% to 25\%, and this is
consistent with the energy spectrum.

\begin{figure}
\rotatebox{-90}{\resizebox{6.0cm}{!}{\includegraphics{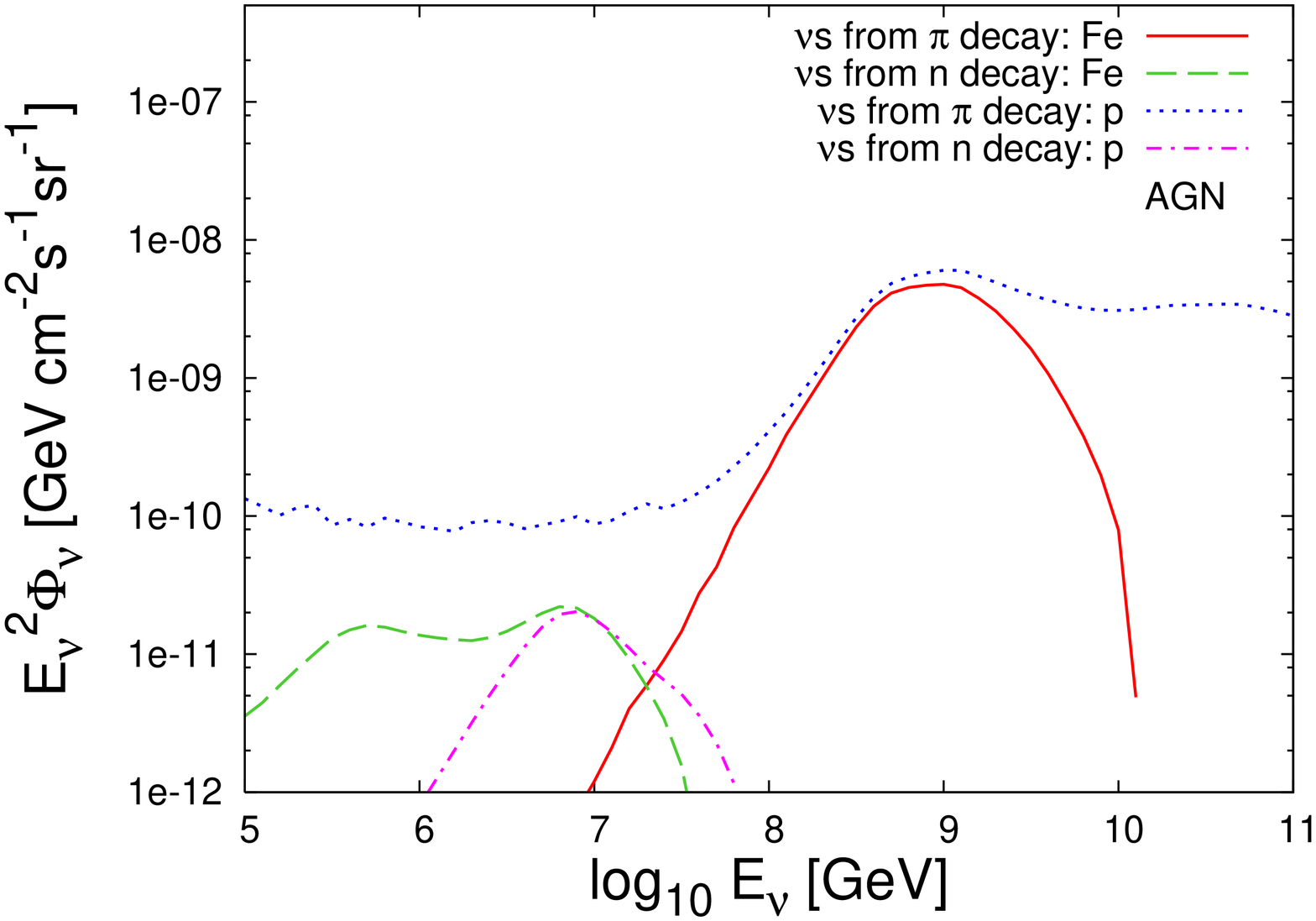}}}
\rotatebox{-90}{\resizebox{6.0cm}{!}{\includegraphics{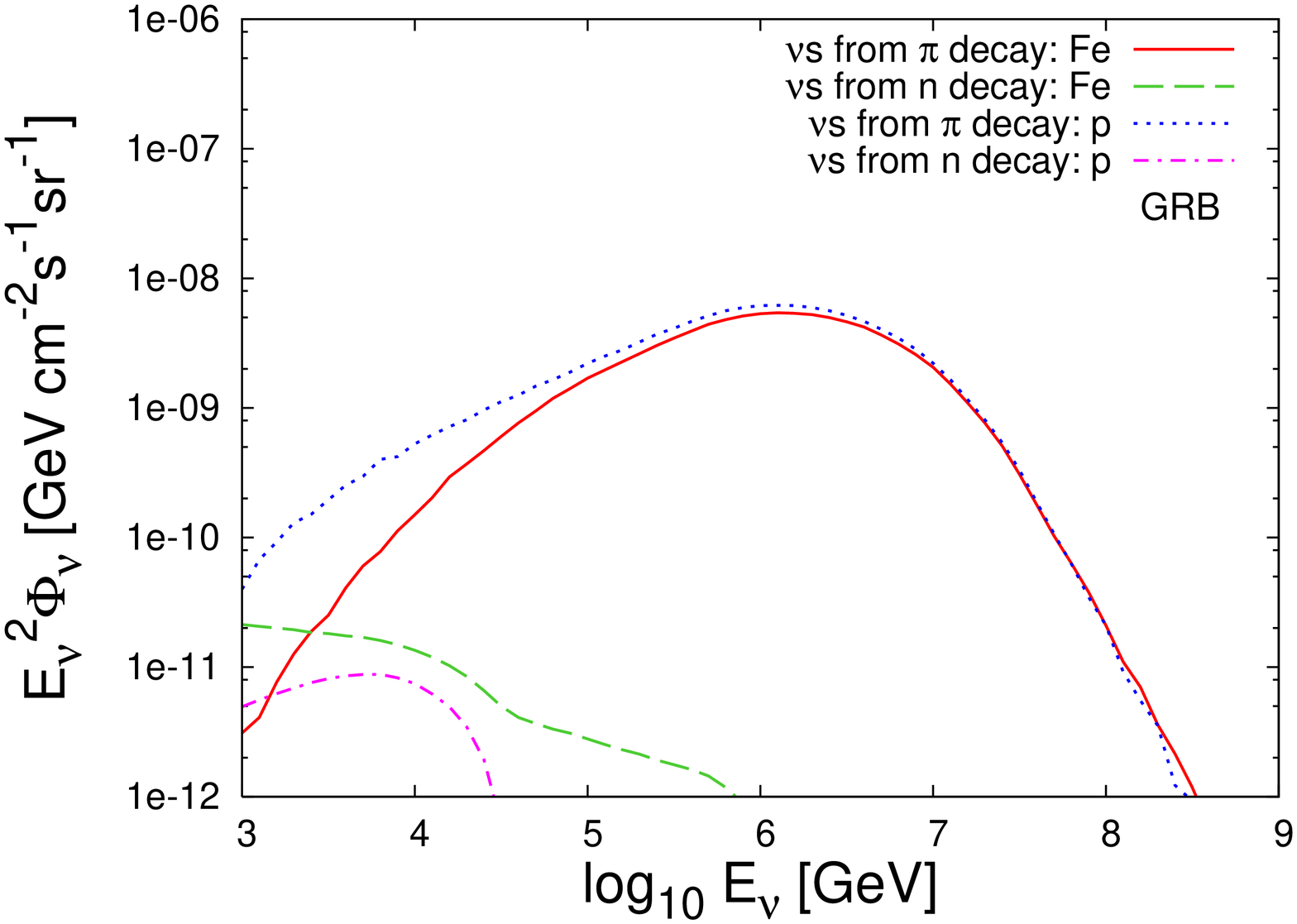}}}
\caption{The neutrino spectra produced by protons and iron nuclei
  being accelerated in AGN (top) and GRBs (bottom). The fluxes have
  been normalized following the Waxman-Bahcall prescription, namely
  assuming that all ultra-high energy particles observed on Earth are
  protons. }
\label{neutrinos}
\end{figure}

By performing a Monte Carlo simulation for cosmic ray protons and
nuclei propagating through the sources considered, the neutrino flux
produced in each of the three case examples can now be calculated. The
results shown in Fig.~\ref{neutrinos} are obtained assuming a
cosmological distribution of sources which accelerate cosmic rays with
a spectrum $\propto E^{-2.4}$. Using the emissivity of AGN, the
corresponding peak neutrino flux is,
\begin{eqnarray}
\Phi_{\nu}(E_\nu) & = & 2.3 
\times 10^{-5}\,\left(1- e^{-\epsilon_\pi/K_p}\right)\nonumber \\
 & \times &    E_\nu^{-2.4} \,
 {\rm GeV}^{-1}\, {\rm cm}^{-2}\,{\rm s}^{-1}\,{\rm sr}^{-1} \,\, ,
\end{eqnarray}
where $E_\nu$ is in GeV and we have normalized to a cosmic ray
production rate of:
\begin{equation}
\dot \epsilon_{\rm CR}^{[10^{18.7},\, 10^{22.0}]} = 2.2 \times
10^{44}\,\rm{erg}\,\rm{Mpc}^{-3} \rm{yr}^{-1}\,\,, 
\end{equation}
as indicated by Auger data. In the case of GRBs, because of the strong
magnetic fields in the plasma, the neutrinos are created by parent
protons with energies below $10^{18}~{\rm eV}$. Following
Ref.~\cite{Berezinsky:2002nc}, we assume in this case a break in the
proton injection spectrum ($\propto E^{-2}$ below $10^{18}~{\rm eV}$),
so that the corresponding peak neutrino flux saturates the
Waxman-Bahcall bound, yielding
\begin{eqnarray}
\Phi_{\nu}(E_\nu) & = & 1.0 
\times 10^{-8}\,\left(1- e^{-\epsilon_\pi/K_p}\right)\nonumber \\
 & \times &    E_\nu^{-2} \,
 {\rm GeV}^{-1}\, {\rm cm}^{-2}\,{\rm s}^{-1}\,{\rm sr}^{-1} \,\, .
\end{eqnarray}
Note that this normalization differs from that in Eq.~(\ref{wbproton})
because of the different energy range of the source injection spectrum
required to accomodate the Auger data~\cite{normalization}.

We do not show the neutrino flux expected from starburst galaxies
because it is negligible compared to that from AGN or GRBs. The
interaction length of protons in starburst galaxies is $\sim 500$
times the source size or larger at all energies (see
Fig.~\ref{ratesS}), leading to $\epsilon_{\pi} \alt 10^{-3}$, while
AGN and GRBs have $\epsilon_{\pi} \gtrsim 1$. Thus the uncertainties in
the modelling of starburst galaxies are immmaterial in this context.

\begin{figure}
\rotatebox{-90}{\resizebox{6.0cm}{!}{\includegraphics{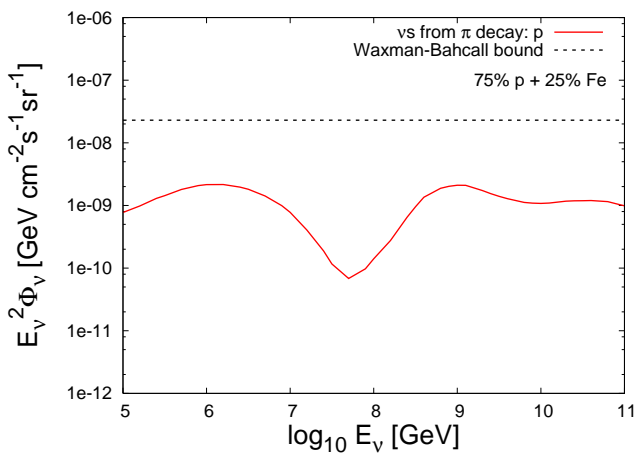}}}
\caption{The neutrino spectrum produced by protons undergoing
  acceleration in AGN and GRBs which are assumed to contribute equally
  to the cosmic ray protons observed at Earth. The Waxman-Bahcall
  bound, as obtained from Eq.~(\ref{wbproton}) by setting
  $\epsilon_\pi = \xi_Z = 1,$ is also shown for comparison.}
\label{numix}
\end{figure}

The required bi-modal composition at source (75\% protons plus 25\% of
heavy nuclei) does however require that sources such as starburst
galaxies accelerate most of the iron nuclei, with most of the protons
coming from AGN and GRBs. In Fig.~\ref{numix}, we show the sum of the
neutrino fluxes produced by AGN and GRBs assuming they contribute
approximately equally. Not surprisingly, the predicted diffuse flux in
this simple model,
\begin{equation}
E_\nu^2 \Phi_\nu \sim 10^{-9}~{\rm GeV}\, 
{\rm cm}^{-2}\, {\rm s}^{-1} \,{\rm sr}^{-1}\,,
\end{equation}
is just the expectation given in Eq.(\ref{simple}), which was based on
the assumption of complete trapping of charged particles (recall that
$\kappa = 0.25$ and, for single $p\gamma$ collisions, $\epsilon_\pi
\sim 0.2$). Of course a more sophisticated estimate can be made using
our results in Fig.\ref{neutrinos} when we know more about the
relative contributions from the different possible sources to the
overall cosmic ray flux, as well as the relative (possibly energy
dependent) weighting of heavy nuclei with respect to protons. Being
conservative, presently we can only argue that the overall cosmic
neutrino flux should be reduced by about 75\%.

In closing, we stress that the diffuse neutrino flux has an additional
component originating in the energy losses of ultra-high energy cosmic
rays {\em en route} to Earth. The main energy loss process here is
photo-pion production in the CMB, which causes the steepening of the
cosmic ray spectrum beyond $10^{19.7}$~eV. The decay of charged pions
produced in this process results in a diffuse flux of ``cosmogenic''
neutrinos~\cite{Stecker:1978ah}, which is comparable to the fluxes
shown in Fig.~\ref{neutrinos}. If there are heavy nuclei in ultra-high
energy cosmic rays then they will preferentially lose energy through
photo-disintegration rather than photo-pion production, so the
cosmogenic neutrino flux will also be suppressed as has been discussed
elsewhere~\cite{Hooper:2004jc,noteave}.

Equipped with the fluxes shown in Figs.~\ref{neutrinos} and
\ref{numix}, we now proceed to determine the expected event rates in a
next generation neutrino telscope such as IceCube.  We do not consider
the cosmogenic flux so these rates are conservative.

\section{Expected event rates at IceCube}
\label{sensitivity}

The IceCube neutrino telescope is currently under construction at the
South Pole. When complete, it will comprise a cubic-kilometer of
ultra-clear ice about 2 km below the surface, instrumented with long
strings of sensitive photon detectors~\cite{Ahrens:2003ix}. We focus
on this experiment as it is the most advanced of its kind, however the
event rates estimated below will also apply to e.g. the KM3 undersea
experiment being planned for the Mediterranean.

IceCube is designed to observe muon tracks and showers produced by
neutrino charged current (CC) and neutral current (NC) interactions in
and around the instrumented volume. The probability of detecting a
neutrino passing through the detector from its muon track is given by
\begin{equation}
P_{\nu \rightarrow \mu}(E_\nu, \theta_{\rm zenith}) = \sigma_{\nu
N}^{\rm CC}(E_\nu)\, n \, R_\mu(E_\mu, \theta_{\rm zenith}), 
\end{equation}
where $\sigma_{\nu N}^{\rm CC}(E_\nu)$ is the charged current
neutrino-nucleon cross section~\cite{Gandhi:1998ri}, $n$ is the number
density of nucleons in the ice, and the muon range
$R_{\mu}(E_{\mu},\theta_{\rm{zenith}})$ is the average distance
traveled by a muon of energy $E_{\mu}$ before it is degraded below
some threshold energy (taken to be 100 GeV). This quantity depends on
the zenith angle of the incoming neutrino as only quasi-horizontal or
upgoing events can benefit from longer muon ranges. At the energies we
are most concerned with, the majority of muon events will be
quasi-horizontal.

The expected muon event rate is
\begin{equation}
\frac{d{\cal N}_{\mu}}{dt} = \int dE_{\nu}\, 
d\Omega \,\Phi_{\nu_\mu}(E_\nu)\, P_{\nu
\rightarrow \mu} (E_{\nu},\theta_{\rm zenith})\, A_{\rm eff}\, ,
\end{equation}
where $\Phi_{\nu_\mu}(E_\nu) = \Phi_\nu(E_\nu)/3$ is the flux of muon
neutrinos and $A_{\rm{eff}} \approx 1~{\rm km}^2$ is the effective
area of the detector~\cite{Ahrens:2003ix}.  Similarly, the expected
number of shower events is
\begin{equation}
\frac{d{\cal N}_{\rm{S}}}{dt} = \sum_\alpha \int dE_{\nu}\, 
d\Omega \,\Phi_{\nu_\alpha}(E_\nu)\, \sigma_{\nu N}^{\rm CC (NC)}(E_{\nu}) \, 
V_{\rm eff}\, , 
\end{equation}
where $\sigma_{\nu N}^{\rm CC (NC)}$ is the CC (NC) neutrino-nucleon
cross section and $V_{\rm eff}$ is the effective volume for detection
of showers~\cite{Anchordoqui:2005is}. The sum is over $\nu_e$ and
$\nu_\tau$ CC interactions and all NC interactions, and we assume
$\Phi_{\nu_\alpha} (E_\nu) = \Phi_\nu(E_\nu)/3$ where $\alpha = e,
\mu, \tau.$ Electron neutrino, CC induced showers carry all of the
incoming neutrino energy, whereas muon neutrino (CC or NC) induced
showers carry away an energy of $(1-y) \, E_{\nu}$, where $y$ is the
inelasticity of the neutrino interaction. Tau neutrino NC
showers have an energy $(1-y) \, E_{\nu}$, whereas tau neutrinos CC
events with energies below $E_{\nu} \sim$ PeV generate showers with
the full energy of the incoming neutrino. We assume that only showers
with energies greater than 3 TeV can be identified at IceCube.

In Table~1, we show the predicted event rates for the various cosmic
ray accelerators considered earlier. We also show the effect on the
rates if the threshold for both muons and showers is taken to be
100~TeV --- a cut at this higher energy is adequate to eliminate
essentially all background from muon bremsstrahlung radiation near the
detector and from muons produced in cosmic ray showers in the
atmosphere.  Moreover, the steeply falling flux of atmospheric
neutrinos is negligible above this energy, so this cut generates a
very pure sample of extraterrestrial neutrinos. The labels ``protons"
and ``iron" denote the species of particle which are accelerated by
the source (rather than what actually escapes). 

As expected, the event rates from GRBs are similar regardless of
whether protons or iron nuclei are accelerated, as the latter are
almost entirely disintegrated by the surrounding radiation fields.
Values for starburst galaxies are not given in the table as the rates
(from $p \gamma$ interactions) are well below the sensitivity of
IceCube and other next generation neutrino telescopes.

\begin{table}
  \caption{Predicted event rates at IceCube for various sources of high
    energy neutrinos, with the muon energy threshold set to 0.1 (100) TeV 
    and the threshold for showers set to be 3 (100)~TeV. 
    The labels ``protons" and ``iron" denote the type of particle assumed 
    to be accelerated, rather than what actually escapes (iron nuclei do not 
    escape intact from GRBs, in particular).
    The label ``UHECR Best Fit" denotes the case where 75\% of
    the ultra high energy cosmic rays arriving at Earth are protons 
    (assumed to be accelerated equally by GRB and AGN) and 25\% are iron nuclei 
    (assumed to be accelerated by starburst galaxies).}
\begin{center}
\begin{tabular}{|c@{}|c|c@{}|c|c|}
  \hline
 ~~~~~~~~Source~~~~~~~~ & ~~~~~~~~$d{\cal N}_{\mu}/dt$~~~~~~~~ & ~~~~~~~~$d{\cal N}_{\rm{S}}/dt$~~~~~~~~ \\
  \hline
  \hline
  AGN (protons) & 1.2 (0.34) yr$^{-1}$ & 0.45 (0.089) yr$^{-1}$ \\
  \hline
  AGN (iron) & 0.23 (0.13) yr$^{-1}$ &  0.045 (0.037) yr$^{-1}$ \\
  \hline
  \hline
  GRB (protons) & 16. (3.4) yr$^{-1}$ & 6.3 (2.2) yr$^{-1}$ \\
  \hline
  GRB (iron) & 12. (2.8) yr$^{-1}$ &  4.5 (1.9) yr$^{-1}$ \\ 
  \hline
  \hline
  UHECR Best Fit & 6.5 (1.4) yr$^{-1}$ & 2.5 (0.86) yr$^{-1}$ \\
  \hline
  \hline
\end{tabular}
\end{center}
\end{table}

\section{Conclusions}
\label{conclusions}

We have studied the role that heavy nuclei play in the generation of
neutrinos in possible astrophysical sources of high energy cosmic
rays. During acceleration the nuclei may be completely
photo-disintegrated into their constituent nucleons and we find this
indeed happens in GRBs, resulting in the outgoing cosmic rays being
proton dominated. The neutrino flux is then left largely unchanged
from previous estimates which had ignored the possibility of nuclei
being accelerated as well as protons. At the other extreme, sources
such as starburst galaxies hardly disintegrate accelerated nuclei,
enabling such particles to escape and to contribute to the observed
ultra-high energy cosmic ray spectrum, largely without contributing to
the cosmic neutrino flux. In AGN the situation is in between, with
nuclei being fully disintegrated only at the highest energies, so the
neutrino flux is suppressed at lower energies.

The likely possibility of a substantial fraction of nuclei in
ultra-high energy cosmic rays implies therefore a somewhat reduced
expectation for the neutrino flux from their cosmic sources. In
particular, as the spectrum and elongation length in the atmosphere of
ultra-high energy cosmic rays appears to be best fitted by a mixture
of 25\% iron nuclei and 75\% protons, the overall neutrino flux is
reduced somewhat relative to the expectation for an all-proton cosmic
ray spectrum.

As next generation neutrino telescopes such as IceCube begin
operation, it becomes increasingly important to refine the
expectations for detection of high energy neutrinos from the sources
of cosmic rays. We have taken into account recent data on ultra-high
energy cosmic rays to provide updated estimates of the neutrino fluxes
to be expected from various types of possible extragalactic
sources. The actual detection of cosmic neutrinos will in turn thus
provide crucial information on the nature of the long sought sources
of cosmic rays.


\acknowledgments{DH is supported by the US DoE and by NASA grant
  NAG5-10842. SS acknowledges a PPARC Senior Fellowship
  (PPA/C506205/1) and the EU network ``UniverseNet''
  (MRTN-CT-2006-035863). We thank Johannes Knapp for providing the
  simulated values of $X_{\rm max}$ and Soeb Razzaque for pointing out
  a typographical error in the discussion on GRBs.}


\begin{thebibliography}{99}
\bibitem{note} 
 The non-thermal energy release in GRBs is much smaller than that output by AGN.
\bibitem{note94}
  The factor $9/(4 R^2)$ results from calculating $\int d^3\vec r_1\ \
  \int d^3\vec r_2\ \ |\vec r_1 - \vec r_2|^{-2}\ (4\pi R^3/3)^{-2},$
  where $\vec r_1$ is the position of a star and $\vec r_2$ is the
  position of an observer (the position of the reaction), in a region
  of radius $R$ uniformly filled with sources.
\bibitem{normalization} 
  It is important to stress that the Auger data are still at a 
  preliminary stage and the reconstruction procedures are still to be
  finalised. However, even allowing for the systematic uncertainties
  still present, it does appear that at the highest energies fewer events
  are seen than expected from the AGASA analysis.
\bibitem{noteave} A point worth noting at this juncture: If iron
  nuclei are accelerated to very high energies (much higher than the
  energy spectrum has been measured), then disintegration can lead to
  large numbers of protons above the spectrum cutoff. In this case,
  the resulting cosmogenic neutrino flux is not dramatically
  suppressed. On the other hand, if iron nuclei are only largely
  accelerated to around $10^{21}$~eV or less, then the liberated
  protons will only rarely interact with the CMB to produce pions,
  hence the cosmogenic neutrino flux will be significantly reduced.
\end{thebibliography}
\end{document}